\pdfoutput = 1
\documentclass[12pt,a4paper]{article}

\usepackage{ifthen}
\newboolean{pdflatex}
\setboolean{pdflatex}{true}

\newboolean{articletitles}
\setboolean{articletitles}{true}

\newboolean{uprightparticles}
\setboolean{uprightparticles}{false}

\usepackage{microtype}
\usepackage{lineno}  
\usepackage{xspace} 

\usepackage{graphicx}  
\usepackage{color}
\usepackage{colortbl}
\graphicspath{{./figs/}} 

\usepackage{amsmath} 
\usepackage{amssymb}
\usepackage{amsfonts}
\usepackage{upgreek} 

\newcommand*\patchAmsMathEnvironmentForLineno[1]{%
\expandafter\let\csname old#1\expandafter\endcsname\csname #1\endcsname
\expandafter\let\csname oldend#1\expandafter\endcsname\csname
end#1\endcsname
 \renewenvironment{#1}%
   {\linenomath\csname old#1\endcsname}%
   {\csname oldend#1\endcsname\endlinenomath}%
}
\newcommand*\patchBothAmsMathEnvironmentsForLineno[1]{%
  \patchAmsMathEnvironmentForLineno{#1}%
  \patchAmsMathEnvironmentForLineno{#1*}%
}
\AtBeginDocument{%
\patchBothAmsMathEnvironmentsForLineno{equation}%
\patchBothAmsMathEnvironmentsForLineno{align}%
\patchBothAmsMathEnvironmentsForLineno{flalign}%
\patchBothAmsMathEnvironmentsForLineno{alignat}%
\patchBothAmsMathEnvironmentsForLineno{gather}%
\patchBothAmsMathEnvironmentsForLineno{multline}%
}

\usepackage{hyperref}    
\usepackage[all]{hypcap} 

\def\lhcb {\mbox{LHCb}\xspace}
\def\ux85 {\mbox{UX85}\xspace}

\ifthenelse{\boolean{uprightparticles}}%
{

 \def\Pmu         {\ensuremath{\upmu}\xspace}

 \def\Ppi         {\ensuremath{\uppi}\xspace}

 \def\Ppsi        {\ensuremath{\uppsi}\xspace}

 \def\PDelta      {\ensuremath{\Delta}\xspace}                
 \def\PXi      {\ensuremath{\Xi}\xspace}                
 \def\PLambda      {\ensuremath{\Lambda}\xspace}                
 \def\PSigma      {\ensuremath{\Sigma}\xspace}                
 \def\POmega      {\ensuremath{\Omega}\xspace}                
 \def\PUpsilon      {\ensuremath{\Upsilon}\xspace}                
 
 \def\PB      {\ensuremath{\mathrm{B}}\xspace}                
                 
 \def\PD      {\ensuremath{\mathrm{D}}\xspace}

 \def\PJ      {\ensuremath{\mathrm{J}}\xspace}                
 \def\PK      {\ensuremath{\mathrm{K}}\xspace}

 \def\Pb      {\ensuremath{\mathrm{b}}\xspace}                
 \def\Pc      {\ensuremath{\mathrm{c}}\xspace}

 \def\Pi      {\ensuremath{\mathrm{i}}\xspace}

 \def\Ps      {\ensuremath{\mathrm{s}}\xspace}

}
{

 \def\Pmu         {\ensuremath{\mu}\xspace}

 \def\Ppi         {\ensuremath{\pi}\xspace}

 \def\Ppsi        {\ensuremath{\psi}\xspace}                
                 
 \mathchardef\PDelta="7101
 \mathchardef\PXi="7104
 \mathchardef\PLambda="7103
 \mathchardef\PSigma="7106
 \mathchardef\POmega="710A
 \mathchardef\PUpsilon="7107
                 
 \def\PB      {\ensuremath{B}\xspace}                
                 
 \def\PD      {\ensuremath{D}\xspace}

 \def\PJ      {\ensuremath{J}\xspace}                
 \def\PK      {\ensuremath{K}\xspace}

 \def\Pb      {\ensuremath{b}\xspace}                
 \def\Pc      {\ensuremath{c}\xspace}

 \def\Pi      {\ensuremath{i}\xspace}

 \def\Ps      {\ensuremath{s}\xspace}

}

\def\mup        {\ensuremath{\Pmu^+}\xspace}
\def\mun        {\ensuremath{\Pmu^-}\xspace} 

\def\ellell     {\ensuremath{\ell^+ \ell^-}\xspace}

\def\squark    {\ensuremath{\Ps}\xspace}

\def\cquark    {\ensuremath{\Pc}\xspace}

\def\bquark    {\ensuremath{\Pb}\xspace}

\def\pion  {\ensuremath{\Ppi}\xspace}

\def\pip   {\ensuremath{\pion^+}\xspace}
\def\pim   {\ensuremath{\pion^-}\xspace}

\def\kaon  {\ensuremath{\PK}\xspace}
  \def\Kbar  {\kern 0.2em\overline{\kern -0.2em \PK}{}\xspace}

\def\Kz    {\ensuremath{\kaon^0}\xspace}
\def\Kzb   {\ensuremath{\Kbar^0}\xspace}
\def\KzKzb {\ensuremath{\Kz \kern -0.16em \Kzb}\xspace}
\def\Kp    {\ensuremath{\kaon^+}\xspace}
\def\Km    {\ensuremath{\kaon^-}\xspace}

\def\KpKm  {\ensuremath{\Kp \kern -0.16em \Km}\xspace}

  \def\Dbar    {\kern 0.2em\overline{\kern -0.2em \PD}{}\xspace}
\def\D       {\ensuremath{\PD}\xspace}

\def\Dz      {\ensuremath{\D^0}\xspace}
\def\Dzb     {\ensuremath{\Dbar^0}\xspace}
\def\DzDzb   {\ensuremath{\Dz {\kern -0.16em \Dzb}}\xspace}
\def\Dp      {\ensuremath{\D^+}\xspace}
\def\Dm      {\ensuremath{\D^-}\xspace}

\def\DpDm    {\ensuremath{\Dp {\kern -0.16em \Dm}}\xspace}

\def\B       {\ensuremath{\PB}\xspace}
  \def\Bbar    {\kern 0.18em\overline{\kern -0.18em \PB}{}\xspace}

\def\Bu      {\ensuremath{\B^+}\xspace}

\def\Bp      {\ensuremath{\Bu}\xspace}

\def\Bd      {\ensuremath{\B^0}\xspace}
\def\Bs      {\ensuremath{\B^0_\squark}\xspace}

\def\jpsi     {\ensuremath{{\PJ\mskip -3mu/\mskip -2mu\Ppsi\mskip 2mu}}\xspace}
\def\psitwos  {\ensuremath{\Ppsi{(2S)}}\xspace}

  \def\Y#1S{\ensuremath{\PUpsilon{(#1S)}}\xspace}

\def\Lbar {\ensuremath{\kern 0.1em\overline{\kern -0.1em\PLambda}}\xspace}

\def\BF         {{\ensuremath{\cal B}\xspace}}

\def\BR         {\BF}
\newcommand{\decay}[2]{\ensuremath{#1\!\to #2}\xspace}         

\def\to                 {\ensuremath{\rightarrow}\xspace}

\def\AT#1     {\ensuremath{A_{\mathrm{T}}^{#1}}\xspace}           

\def\C#1      {\ensuremath{\mathcal{C}_{#1}}\xspace}                       
\def\Cp#1     {\ensuremath{\mathcal{C}_{#1}^{'}}\xspace}                    
\def\Ceff#1   {\ensuremath{\mathcal{C}_{#1}^{\mathrm{(eff)}}}\xspace}        
\def\Cpeff#1  {\ensuremath{\mathcal{C}_{#1}^{'\mathrm{(eff)}}}\xspace}       
\def\Ope#1    {\ensuremath{\mathcal{O}_{#1}}\xspace}                       
\def\Opep#1   {\ensuremath{\mathcal{O}_{#1}^{'}}\xspace}                    

\newcommand{\tev}{\ensuremath{\mathrm{\,Te\kern -0.1em V}}\xspace}
\newcommand{\gev}{\ensuremath{\mathrm{\,Ge\kern -0.1em V}}\xspace}
\newcommand{\mev}{\ensuremath{\mathrm{\,Me\kern -0.1em V}}\xspace}
\newcommand{\kev}{\ensuremath{\mathrm{\,ke\kern -0.1em V}}\xspace}
\newcommand{\ev}{\ensuremath{\mathrm{\,e\kern -0.1em V}}\xspace}
\newcommand{\gevc}{\ensuremath{{\mathrm{\,Ge\kern -0.1em V\!/}c}}\xspace}
\newcommand{\mevc}{\ensuremath{{\mathrm{\,Me\kern -0.1em V\!/}c}}\xspace}
\newcommand{\gevcc}{\ensuremath{{\mathrm{\,Ge\kern -0.1em V\!/}c^2}}\xspace}
\newcommand{\gevgevcccc}{\ensuremath{{\mathrm{\,Ge\kern -0.1em V^2\!/}c^4}}\xspace}
\newcommand{\mevcc}{\ensuremath{{\mathrm{\,Me\kern -0.1em V\!/}c^2}}\xspace}

\def\mum  {\ensuremath{\,\upmu\rm m}\xspace}

\def\sy    {\ensuremath{\sigma_y}\xspace}

\newcommand{\stat}{\ensuremath{\mathrm{(stat.)}}\xspace}
\newcommand{\syst}{\ensuremath{\mathrm{(syst.)}}\xspace}

\def\gsim{{~\raise.15em\hbox{$>$}\kern-.85em
          \lower.35em\hbox{$\sim$}~}\xspace}
\def\lsim{{~\raise.15em\hbox{$<$}\kern-.85em
          \lower.35em\hbox{$\sim$}~}\xspace}

\def\pt         {\mbox{$p_{\rm T}$}\xspace}

\def\evtgen     {\mbox{\textsc{EvtGen}}\xspace}
\def\pythia     {\mbox{\textsc{Pythia}}\xspace}

\def\geant      {\mbox{\textsc{Geant4}}\xspace}

\def\photos     {\mbox{\textsc{Photos}}\xspace}

\def\tell1  {TELL1\xspace}
\def\ukl1   {UKL1\xspace}

\usepackage{cite} 
\usepackage{mciteplus}

\def\jpsik {\decay{\Bp}{\jpsi\Kp}}
\def\jpsikmm {\decay{\Bp}{\jpsi (\to \mup \mun) \Kp}}
\def\jpsipi {\decay{\Bp}{\jpsi\pip}~}
\def\kmumu {\decay{\Bp}{\Kp \mup \mun}}
\def\pimumu {\decay{\Bp}{\pip \mup \mun}}

\def\dmumu {{\decay{b}{d  \mup \mun}}}
\def\dellell {\decay{b}{d \ellell}}
\def\sellell {\decay{b}{s \ellell}}
\def\dy {\decay{b}{d \gamma}}
\def\sy {\decay{b}{s \gamma}}
\def\smumu {\decay{b}{s  \mup \mun}}
\def\sdmumu {\decay{b}{(s,d)  \mup \mun}}

\def\vtdts {$|V_{\text{td}}| / |V_{\text{ts}}|$}
\def\kpipi {\decay{\Bp}{\Kp \pip \pim}}
\def\dkpipi {\decay{\Bp}{ \Dzb (\to \Kp \pim) \pip}}
 \def\kpifrac {$R$}
\def\photos     {\mbox{\textsc{Photos}}\xspace}

\newcommand{\Jpsimumu}{\ensuremath{J/\psi\to \mu^+\mu^-}\xspace}

\begin{document}

\renewcommand{\thefootnote}{\fnsymbol{footnote}}
\setcounter{footnote}{1}

\begin{titlepage}
\pagenumbering{roman}

\vspace*{-1.5cm}
\centerline{\large EUROPEAN ORGANIZATION FOR NUCLEAR RESEARCH (CERN)}
\vspace*{1.5cm}
\hspace*{-0.5cm}
\begin{tabular*}{\linewidth}{lc@{\extracolsep{\fill}}r}
\ifthenelse{\boolean{pdflatex}}
{\vspace*{-2.7cm}\mbox{\!\!\!\includegraphics[width=.14\textwidth]{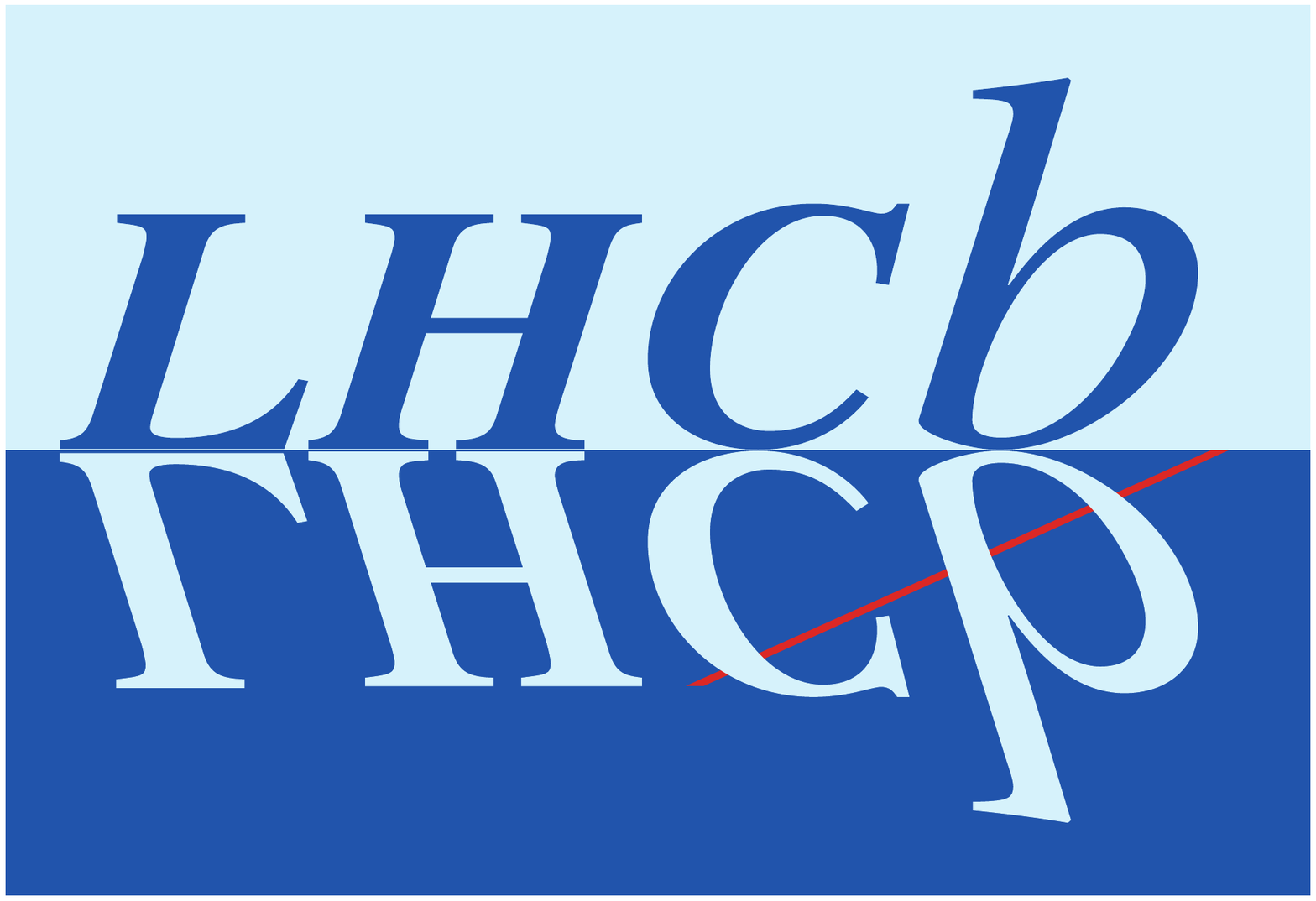}} & &}%
{\vspace*{-1.2cm}\mbox{\!\!\!\includegraphics[width=.12\textwidth]{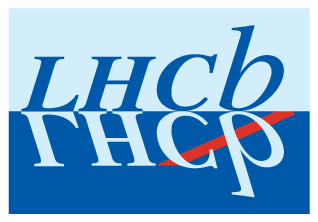}} & &}%
\\
 & & CERN-PH-EP-2012-284 \\
 & & LHCb-PAPER-2012-020 \\
 & & October 9, 2012 \\
 & & \\
\end{tabular*}

\vspace*{4.0cm}

{\bf\boldmath\huge
\begin{center}
First observation of the decay $\Bp \to \pip \mu^+\mu^-$
\end{center}
}

\vspace*{2.0cm}

\begin{center}
The LHCb collaboration\footnote{Authors are listed on the following pages.}
\end{center}

\vspace{\fill}

\begin{abstract}
  \noindent
A discovery of the rare decay $\Bp \to \pip \mu^+\mu^-$ is presented. 
This decay is observed for the first time, with 5.2 $\sigma$ significance.
The observation is made using $pp$ collision data, corresponding to an integrated luminosity of 1.0~fb$^{-1}$, collected with the LHCb detector.
The measured branching fraction is (2.3~$\pm$~0.6~(stat.)~$\pm$~0.1~(syst.))$\times 10^{-8}$, and the ratio of 
the \pimumu and \mbox{\kmumu} branching fractions is measured to be 0.053~$\pm$~0.014~(stat.)~$\pm$~0.001~(syst.).
\end{abstract}

\vspace*{2.0cm}

\begin{center}
  Published in the Journal of High Energy Physics
\end{center}

\vspace{\fill}

\end{titlepage}

\newpage
\setcounter{page}{2}
\mbox{~}
\newpage

\centerline{\large\bf LHCb collaboration}
\begin{flushleft}
\small
R.~Aaij$^{38}$, 
C.~Abellan~Beteta$^{33,n}$, 
A.~Adametz$^{11}$, 
B.~Adeva$^{34}$, 
M.~Adinolfi$^{43}$, 
C.~Adrover$^{6}$, 
A.~Affolder$^{49}$, 
Z.~Ajaltouni$^{5}$, 
J.~Albrecht$^{35}$, 
F.~Alessio$^{35}$
% ...

{\footnotesize \it
$ ^{1}$Centro Brasileiro de Pesquisas F\'{i}sicas (CBPF), Rio de Janeiro, Brazil\\
% ...
$ ^{5}$Clermont Universit\'{e}, Universit\'{e} Blaise Pascal, CNRS/IN2P3, LPC, Clermont-Ferrand, France\\
$ ^{6}$CPPM, Aix-Marseille Universit\'{e}, CNRS/IN2P3, Marseille, France\\
% ...
$ ^{11}$Physikalisches Institut, Ruprecht-Karls-Universit\"{a}t Heidelberg, Heidelberg, Germany\\
% ...
$ ^{33}$Universitat de Barcelona, Barcelona, Spain\\
$ ^{34}$Universidad de Santiago de Compostela, Santiago de Compostela, Spain\\
$ ^{35}$European Organization for Nuclear Research (CERN), Geneva, Switzerland\\
% ...
$ ^{38}$Nikhef National Institute for Subatomic Physics, Amsterdam, The Netherlands\\
% ...
$ ^{43}$H.H. Wills Physics Laboratory, University of Bristol, Bristol, United Kingdom\\
% ...
$ ^{49}$Oliver Lodge Laboratory, University of Liverpool, Liverpool, United Kingdom\\
% ...
$ ^{n}$LIFAELS, La Salle, Universitat Ramon Llull, Barcelona, Spain\\
% ...
}
\end{flushleft}

\cleardoublepage

\renewcommand{\thefootnote}{\arabic{footnote}}
\setcounter{footnote}{0}

\pagestyle{plain}
\setcounter{page}{1}
\pagenumbering{arabic}

\linenumbers


\section{Introduction}
\label{sec:Introduction}

The ratio of Cabibbo-Kobayshi-Maskawa matrix~\cite{CKM} elements~\vtdts~has been measured in \B mixing processes, where it is probed in box diagrams through the ratio of \Bd and \Bs mixing frequencies~\cite{LHCb-PAPER-2011-010,PhysRevLett.97.242003,Bazavov:2012zs,Amhis:2012bh}. The ratio of these matrix elements has also been measured using the ratio of branching fractions of \sy and \dy~decays, where radiative penguin diagrams mediate the transition~\cite{delAmoSanchez:2010ae,Abe:2005rj,Aubert:2006pu}. These measurements of \vtdts~are consistent, within the (dominant) $\sim$10\% uncertainty on the determination from radiative decays. 
The decays \smumu and \dmumu~offer an alternative way of measuring \vtdts~which is sensitive to different classes of operators than the radiative decay modes~\cite{Hurth:2010tk}.  These \mbox{\sdmumu} transitions are flavour-changing neutral current processes which are forbidden at tree level in the Standard Model (SM). In the SM, the branching fractions for \dellell transitions are suppressed relative to \mbox{\sellell} processes by the ratio {$|V_{\text{td}}|^{2} / |V_{\text{ts}}|^{2}$}. This suppression does not necessarily apply to models beyond the SM, and \pimumu decays\footnote{Charge conjugation is implicit throughout this paper.} may be more sensitive to the effect of new particles than \kmumu decays.
In the SM, the ratio of branching fractions for these exclusive modes
\begin{equation}
R \equiv {\BR(\pimumu)}~/~{\BR(\kmumu)}
\end{equation}
is given by 
$R = V^{2}f^{2}$, where $V = |V_{\text{td}}| / |V_{\text{ts}}|$ and $f$ is the ratio of the relevant form factors and Wilson coefficients, integrated over the relevant phase space.
A difference between the measured value of $R$ and $V^{2}f^{2}$ would indicate a deviation from the minimal flavour violation hypothesis~\cite{MFV, Feldmann:2006jk}, and would rule out certain approximate flavour symmetry models~\cite{lessMFV}.

No \mbox{\dellell} transitions have previously been detected, 
and the observation of the \mbox{\pimumu} decay would therefore be the first time such a process has been seen. The predicted SM branching fraction for \mbox{\pimumu} is \mbox{(2.0 $\pm$ 0.2)$\times 10^{-8}$}~\cite{Wang:2007sp}. The most stringent limit to date is \BR{({\decay{\Bp}{\pip \mup\mun}})} $< 6.9 \times 10^{-8}$ at 90\% confidence level~\cite{Wei:2008nv}. The analogous \sellell decay, \kmumu, has been observed with a branching fraction of (4.36$~\pm~$0.15$~\pm~$0.18)~$\times~ 10^{-7}$~\cite{blake}.

This paper describes the search for the \pimumu decay using $pp$ collision data, corresponding to an integrated luminosity of 1.0~fb$^{-1}$, collected with the LHCb detector. 
The \pimumu branching fraction is measured with respect to that of \jpsikmm, and the ratio of \pimumu and \kmumu branching fractions is also determined.


The \lhcb detector~\cite{Alves:2008zz} is a single-arm forward
spectrometer covering the pseudo-rapidity range $2<\eta <5$. The experiment is designed
for the study of particles containing \bquark or \cquark quarks. The
apparatus includes a high precision tracking system, consisting of a
silicon-strip vertex detector surrounding the $pp$ interaction region, and
a large-area silicon-strip detector located upstream of a dipole
magnet. The dipole magnet has a bending power of about $4{\rm\,Tm}$. Three stations
of silicon-strip detectors and straw drift-tubes are placed
downstream of the magnet. The combined tracking system has a momentum resolution
$\Delta p/p$ that varies from 0.4\% at momenta of 5\gevc, to 0.6\% at 100\gevc. The tracking system gives 
an impact parameter resolution of 20\mum for tracks with a 
high transverse momentum (\pt). Charged hadrons are identified using two
ring-imaging Cherenkov detectors. Photon, electron and hadron
candidates are identified by a calorimeter system consisting of
scintillating-pad and preshower detectors, an electromagnetic
calorimeter and a hadronic calorimeter. Muons are identified by a 
system composed of alternating layers of iron and either multi-wire proportional chambers or triple gaseous electron multipliers.

In the present analysis, events are first required to have passed a hardware trigger which selects high-\pt single muons or dimuons. In the first stage of the subsequent software trigger, a single high impact parameter and high-\pt track is required. 
In the second stage of the software trigger, events are reconstructed and then selected for storage based on either the (partially) reconstructed $B$ candidate or the dimuon candidate~\cite{LHCb-PUB-2011-016,Aaij:1384386}.

To produce simulated samples of signal and background decays, $pp$ collisions are generated using
\pythia~6.4~\cite{Sjostrand:2006za} with a specific \lhcb
configuration~\cite{LHCb-PROC-2010-056}.  Decays of hadronic particles
are described by the \evtgen package~\cite{Lange:2001uf} in which final state
radiation is generated using \photos~\cite{Golonka:2005pn}. The
interaction of the generated particles with the detector and the detector 
response are implemented using the \geant
toolkit~\cite{Allison:2006ve, *Agostinelli:2002hh}, as described in
Ref.~\cite{LHCb-PROC-2011-006}.

The small branching fractions of the \pimumu and \kmumu signal decays necessitate good control of the backgrounds and the use of suitably constrained models to fit the invariant-mass distributions. The decay \jpsikmm (hereafter denoted \jpsik) is used to extract both the shape of the signal mass peaks and, in the \pimumu case, the invariant mass distribution of the misidentified \kmumu events. These misidentified \kmumu events form the main residual background after the application of the selection requirements. 

\section{Event selection}
\label{sec:selection}

The \mbox{\pimumu} and \mbox{\kmumu} candidates are selected by combining pairs of oppositely charged muons with a charged pion or kaon.
The selection includes requirements on the impact parameters of the final-state particles and \B candidate, the vertex quality and displacement of the \B candidate, particle identification (PID) requirements 
on the muons and 
a requirement that the \B candidate momentum vector points to one of the primary vertices in the event.
 The rate of events containing more than one reconstructed candidate is 1 in $\sim$20,000 for \jpsik. No restriction is therefore placed on the number of candidates per event.

The pion identification requirements select a sample of pions with an efficiency of $\sim$70\% and a kaon rejection of 99\%. 
The kaon identification requirements allow the selection of a mutually exclusive sample with similar efficiencies.
The muon identification requirements have an efficiency of $\sim$80\%, with a pion rejection of $\sim$99.5\%. 
The PID requirements have a momentum dependent efficiency which is measured from data, in bins of
momentum, pseudorapidity and track multiplicity. The efficiency of the
hadron PID requirements is measured from a sample of $D^{*+} \to (D^{0} \to \Km \pip) \pip$ candidates that allows the hadrons to be unambiguously identified based on their 
kinematics. The muon PID efficiencies are measured using \jpsik candidates, using a tag and probe method.

The \jpsi and \psitwos resonances, where ${\PJ\mskip -3mu/\mskip -2mu\Ppsi\mskip 2mu}, {\Ppsi{(2S)}} \to \mup \mun$, are excluded using a veto on the dimuon mass. This veto has a total width of 200 (150) \mevcc around the nominal \jpsi ($\psi (2S)$) mass~\cite{PDG2012}, and takes into account the radiative tail of these decays. Candidates where the dimuon mass is poorly measured have a correlated mismeasurement in the $ h \mu \mu$ mass. The veto therefore includes a component which shifts with $ h \mu \mu$ mass to exclude such candidates.
 Several other backgrounds are considered: combinatorial backgrounds, where the particles selected do not originate from a single decay; peaking backgrounds, where a single decay is selected but with one or more particles misidentified; and partially reconstructed backgrounds, where one or more final-state particles from a \B decay are not reconstructed. These backgrounds are each described below.

\subsection{Combinatorial backgrounds}

A boosted decision tree (BDT)~\cite{Breiman} which employs the AdaBoost algorithm\cite{AdaBoost} is used to separate signal candidates from the combinatorial background.
Kinematic and geometric properties of the \Bp candidate and final state particles, \mbox{\Bp candidate} 
vertex quality and final state particle track quality are input variables to the BDT.

\begin{figure}
  \centering
   \includegraphics[width=0.6\textwidth] {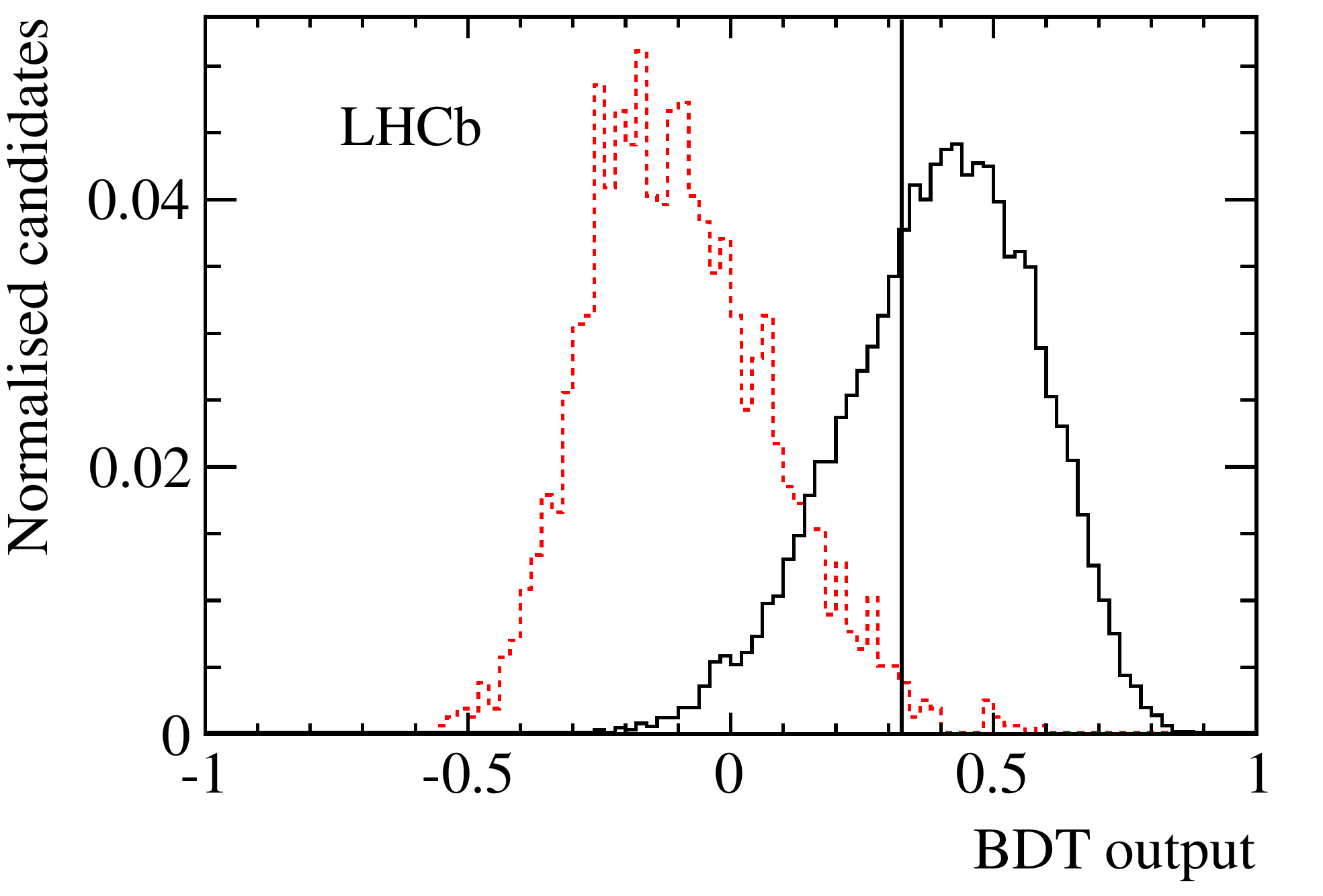}

  \caption{BDT output distribution for simulated \pimumu events (black solid line) and candidates taken from the mass sidebands in the data (red dotted line). Both distributions are normalised to unit area. The vertical line indicates the chosen cut value of 0.325.}
   \label{fig:bdt}
\end{figure}

The BDT is trained on a simulated \pimumu signal sample, and a background sample taken from sidebands in the \pimumu and \kmumu invariant mass distributions. These invariant masses are denoted $M_{\pip \mup \mun}$ and $M_{\Kp \mup \mun}$, respectively. The background sample consists of 20\% of the candidates with $M_{\pip \mup \mun}$ or $M_{\Kp \mup \mun}$~$>$~5500~\mevcc. 
This sample is not used for any of the subsequent analysis.
Signal candidates are required to have a BDT output which exceeds a set value. This value is determined by simulating an ensemble of datasets with the expected signal and background yields, and choosing the cut value which gives the best statistical significance for the \pimumu signal yield. The same method is used to select the optimal set of PID requirements. The BDT output distribution for simulated \pimumu events and for mass sideband candidates 
is shown in Fig.~\ref{fig:bdt}. A cut on the BDT output~$>$~0.325 reduces the expected combinatorial background from \mbox{652 $\pm$ 11} to \mbox{9 $\pm$ 2} candidates in a $\pm$60~\mevcc window around the nominal \B mass, while retaining 68\% of signal events. Assuming the SM branching fraction and the single event sensitivity defined in Sect.~\ref{sec:norm}, \mbox{21 $\pm$ 3 \pimumu} signal events are expected in the data sample. 

\subsection{Peaking and partially reconstructed backgrounds}

Backgrounds from fully reconstructed \Bp decays with one or more misidentified particles have a peaking mass structure. 
After applying the PID requirements, the fraction of \mbox{\kmumu} candidates misidentified as \mbox{\pimumu} is 0.9\%, giving a residual background expectation of 6.2 $\pm$ 0.3 candidates. This expectation is computed
by weighting \kmumu candidates, isolated using a kaon PID requirement,
according to the PID efficiency obtained from the $D^{*+}$ calibration sample. 
The only other decay found to give a significant peaking background in the search for \pimumu is $B^{+} \rightarrow \pi^{+} \pi^{+} \pi^{-}$, where both a \pip and a \pim are 
misidentified as muons. For \kmumu decays, the only significant peaking background is \kpipi, which includes the contribution from \mbox{\dkpipi}. 
The expected background levels from \mbox{$B^{+} \rightarrow \pi^{+} \pi^{+} \pi^{-}$ (\kpipi)} decays are computed to be \mbox{0.39 $\pm$ 0.04} (\mbox{1.56 $\pm$ 0.16}) residual background candidates, using simulated events.

Backgrounds from decays that have one or more final state particles which are not reconstructed have a mass below the nominal $B$ mass, and do not extend into the signal window. However, in the \pimumu case, these backgrounds overlap with the misidentified \kmumu component described above, and must therefore be included in the fit. In the \kmumu case such partially reconstructed backgrounds are negligible.

\subsection{Control channels}
The \jpsik and \kmumu decay candidates are isolated by replacing the pion PID criteria with a requirement to select kaons. In addition, instead of the dimuon mass vetoes described above, the \jpsik candidates are required to have dimuon mass within $\pm$50 \mevcc of the nominal \jpsi mass (the \jpsi mass resolution is 14.5~\mevcc). The remainder of the selection is the same as that used for \pimumu. 
This minimises the systematic uncertainty on the ratio of branching fractions, although the selection is considerably tighter than that which would give the lowest statistical uncertainty on the \kmumu event yield. 
The ${B^+\to (J\!/\!\psi \to \mup \mun) \pip}$ candidates (denoted ${B^+\to J\!/\!\psi \pip}$), which are discussed below, are selected using the same BDT, the pion PID criteria, and the above window on the dimuon invariant mass. 
There is no significant peaking background for \jpsik decays. For \jpsipi decays the only significant peaking background is misidentified \jpsik events.

\section{Signal yield determination}
\label{sec:fit}
The \pimumu, \kmumu and \jpsik yields are determined from a simultaneous
unbinned maximum likelihood fit to four invariant mass distributions which contain: 
\begin{enumerate}
 \item{Reconstructed \jpsik candidates;}
 \item{Reconstructed \jpsik candidates, with the kaon attributed to have the pion mass;}
 \item{Reconstructed \pimumu candidates; and}
 \item{Reconstructed \kmumu candidates.}
\end{enumerate}

The signal probability density functions (PDFs) for the \pimumu, \kmumu, and \jpsik decay modes
 are modelled with the sum of two Gaussian functions. The PDFs for all of these decay modes share the same mean, widths and fraction of the total PDF between the two Gaussians. 
The \pimumu PDF is adjusted for the difference between the widths of the \pimumu and \jpsik distributions, which is observed to be at the percent level in simulation.
The peaking backgrounds described in Sect.~2.2 are taken into account in the fit by including PDFs with shapes determined from simulation.
 The combinatorial backgrounds are modelled with a single exponential PDF, with the exponent allowed to vary independently for each distribution.
The partially reconstructed candidates are modelled using a PDF consisting of an exponential distribution cut-off at a threshold mass, with the transition smeared by the experimental resolution. The shape parameters are again allowed to vary independently for each distribution.
The misidentified \jpsik candidates are modelled with a Crystal Ball function~\cite{Skwarnicki:1986xj}, as it describes the shape well. 
In order to describe the relevant background components for \pimumu, the fit is performed in the mass range 4900 $<$ $M_{\pip \mup \mun}$ $<$ 7000 \mevcc. To avoid fitting the partially reconstructed background for \kmumu, which is irrelevant for the analysis, the fit is performed in the mass range \mbox{5170 $<$ $M_{\Kp \mup \mun}$ $<$ 7000 \mevcc}.

\subsection{\boldmath{Reconstructed \jpsik candidates}}
The reconstructed \jpsik candidates are shown in the $M_{\Kp \mup \mun}$ distribution in Fig.~\ref{fig:both}(a). 
The fitted \mbox{\jpsik} yield is \mbox{106,230 $\pm$ 330}.  This large event yield determines the lineshape for the \mbox{\pimumu} and \kmumu signal distributions, and provides the normalisation for the \pimumu branching fraction.

\subsection{Reconstructed \boldmath{\jpsik candidates with the pion mass hypothesis}}
The \jpsik candidates reconstructed under the pion mass hypothesis provide the lineshape for the misidentified \kmumu candidates that are a background to the \pimumu signal. 
The equivalent background from \pimumu in the \kmumu sample is negligible.

\begin{figure}
  \centering
   \includegraphics[width=0.49\textwidth] {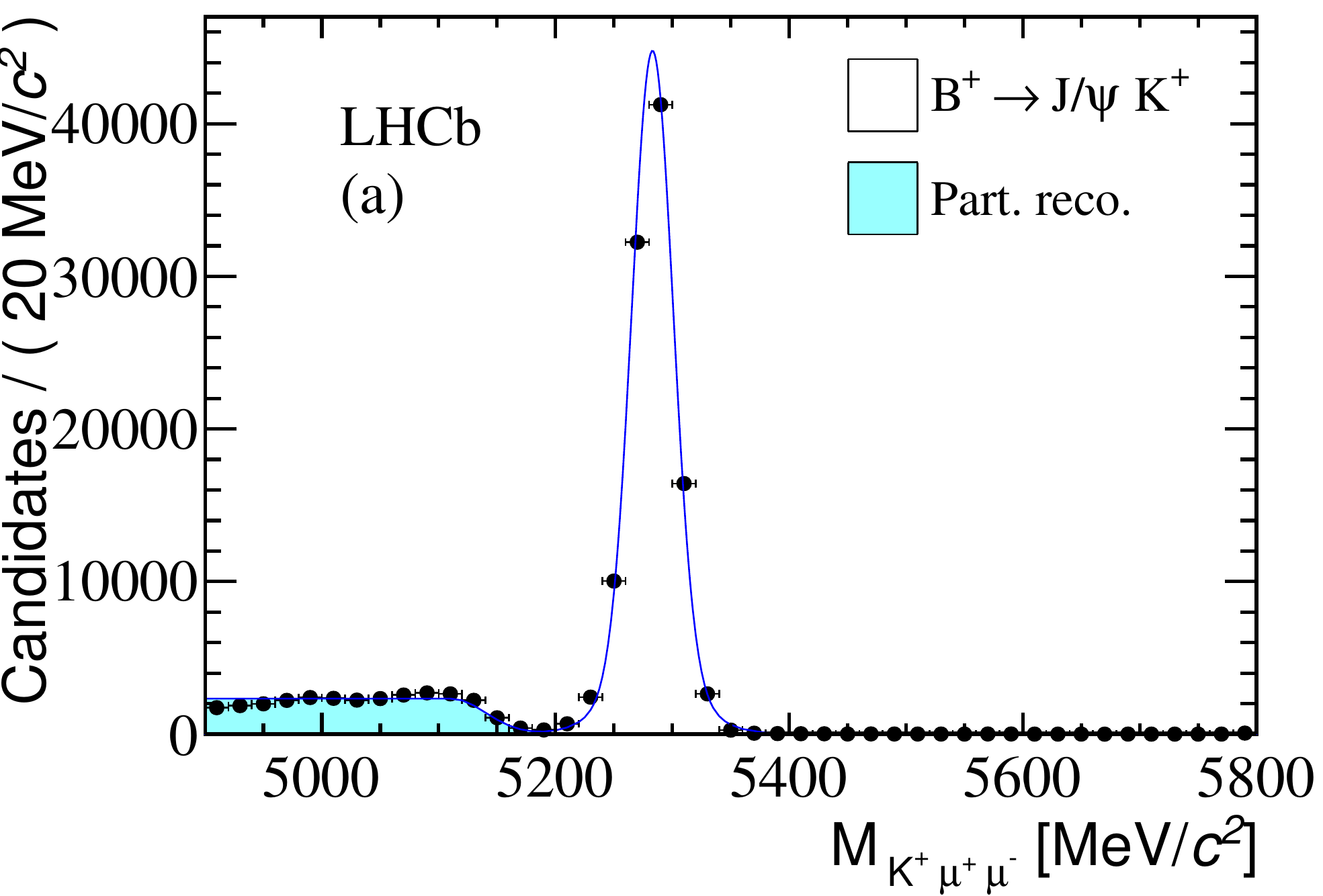}
            \label{fig:pionfit}
      \includegraphics[width=0.49\textwidth] {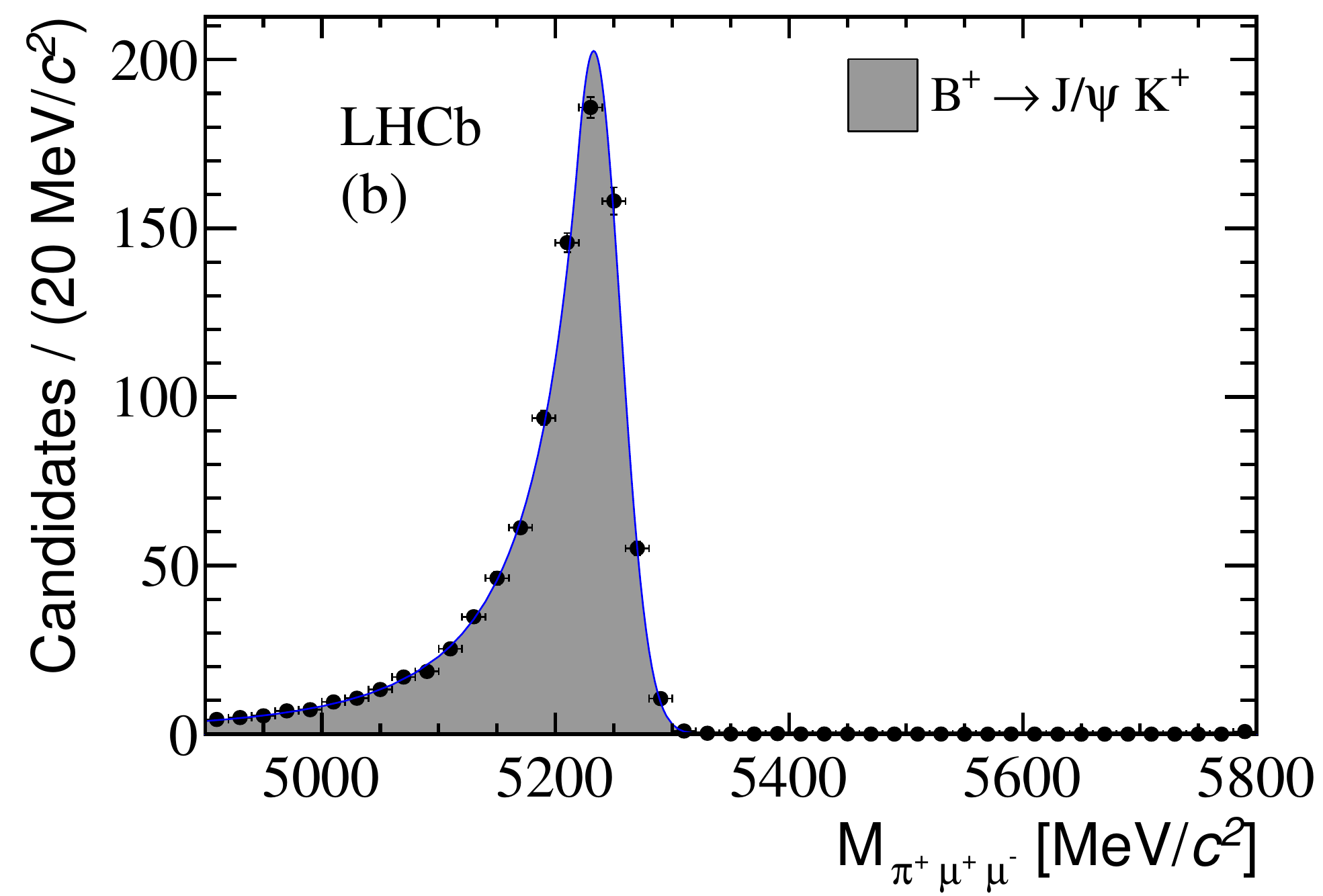}
               \label{fig:kaonfit}
   
   		\label{fig:both}

  \caption{Invariant mass distribution for \jpsik candidates under the (a) $\Kp \mup \mun$ and (b) $\pip \mup \mun$ mass hypotheses with the fit projections overlaid. In the legend, ``part. reco'' refers to partially reconstructed background.
  The fit models are described in the text.}
\end{figure}

The PID requirements used in the selection have a momentum dependent efficiency and therefore
change the mass distribution of any backgrounds with candidates that have misidentified particles. 
In order to correct for this effect, the \jpsik candidates are reweighted according to the PID efficiencies derived from data, as described in Sect.~2.2.
This adjusts the \jpsik invariant mass distribution
to remove the effect of the
kaon PID requirement used to isolate \jpsik, and to reproduce
the effect of the pion PID requirement used to isolate \pimumu.
In addition, there is a difference in the lineshapes of the \jpsik
and \kmumu invariant mass distributions under the pion mass hypothesis. This effect arises from the
differences between the two decay modes' dimuon energy and hadron
momentum spectra, and is therefore corrected by reweighting \jpsik candidates in
terms of these variables.
The $M_{\pip \mup \mun}$ distribution after both weighting procedures have been applied is shown in Fig.~\ref{fig:both}(b).

\subsection{Reconstructed \boldmath{\pimumu} and \boldmath{\kmumu} candidates}

The yield of misidentified \kmumu candidates in the \pimumu
invariant mass distribution is constrained to the expectation
given in Sect.~2.2. Performing the fit without this constraint gives a yield of \mbox{5.6 $\pm$ 6.4} misidentified \mbox{\kmumu} candidates. 
The yields for the peaking background components are constrained to the expectations given in Sect.~2.2.
For both the $M_{\pip \mup \mun}$ and $M_{\Kp \mup \mun}$ distributions, the exponential PDF used to model the combinatorial background has a step in the normalisation at 5500~\mevcc to account for the data used for training the BDT.

\begin{figure}
  \centering
   \includegraphics[width=0.49\textwidth] {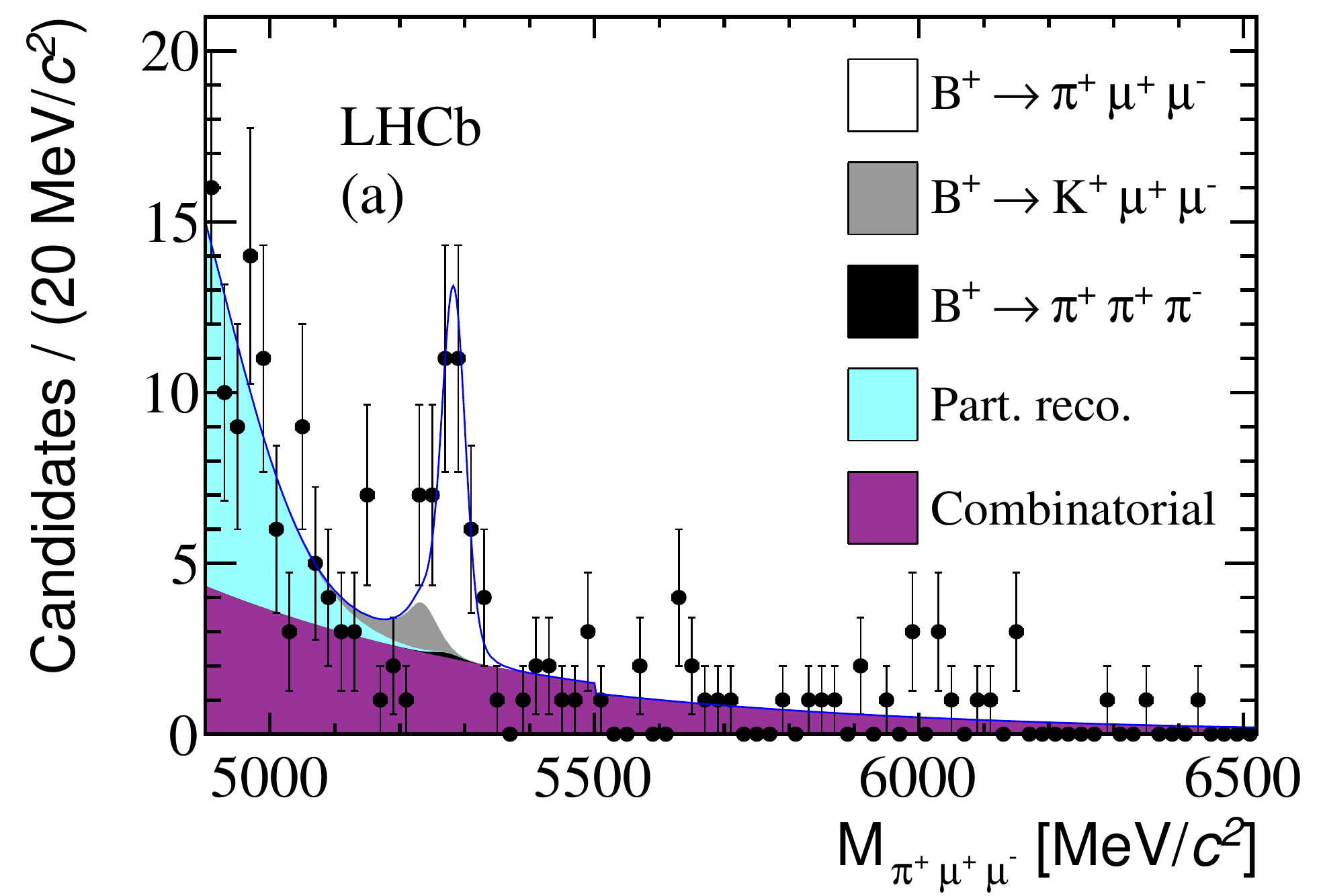}
   \includegraphics[width=0.49\textwidth] {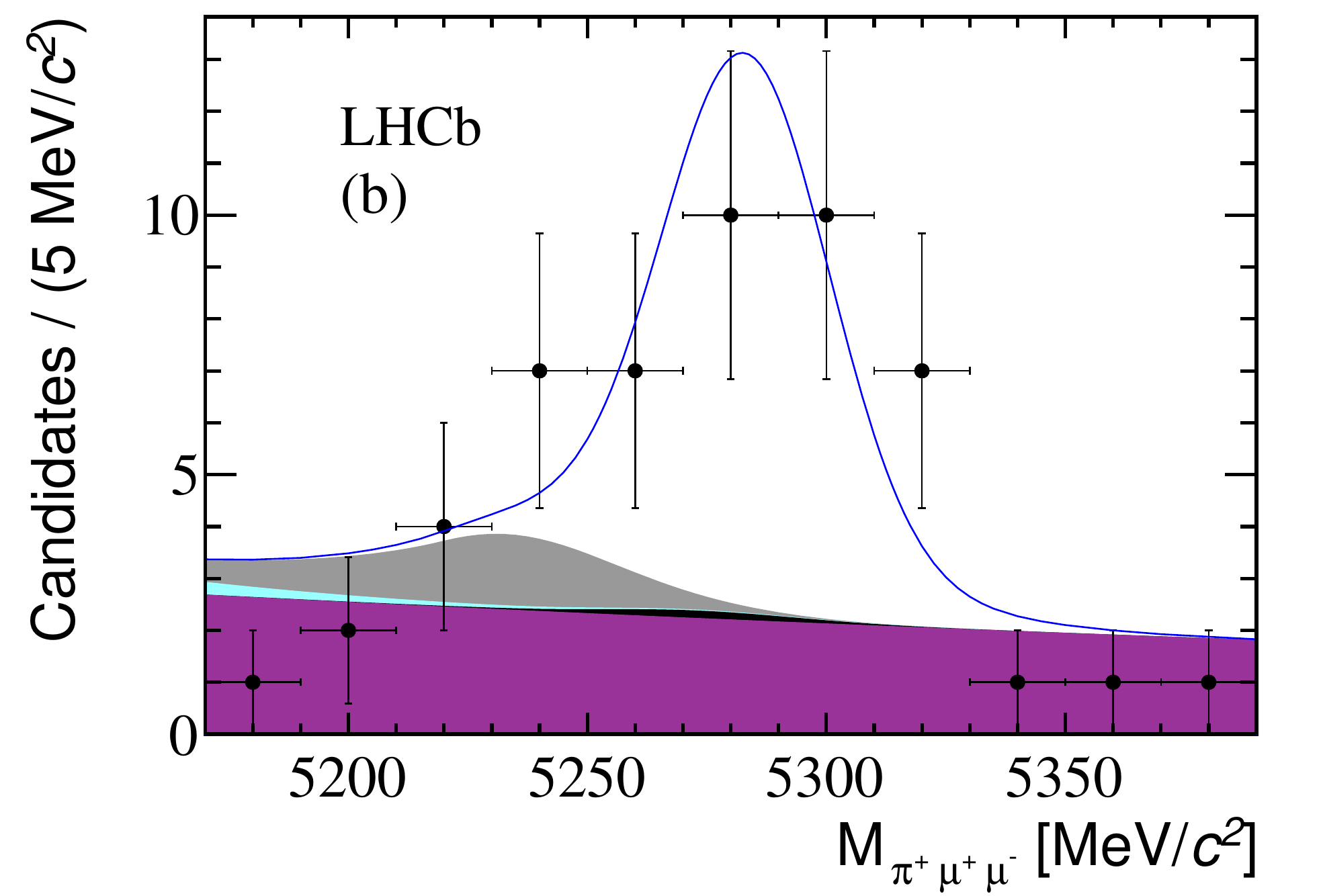}

  \caption{Invariant mass distribution of \pimumu candidates with the fit projection overlaid (a) in the full mass range and (b) in the region around the \B mass. In the legend, ``part. reco.'' and ``combinatorial'' refer to partially reconstructed and combinatorial backgrounds respectively. 
The discontinuity at 5500 \mevcc is due to the removal of data used for training the BDT.}
   \label{fig:pimumu}
\end{figure}

\begin{figure}
  \centering
   \includegraphics[width=0.49\textwidth] {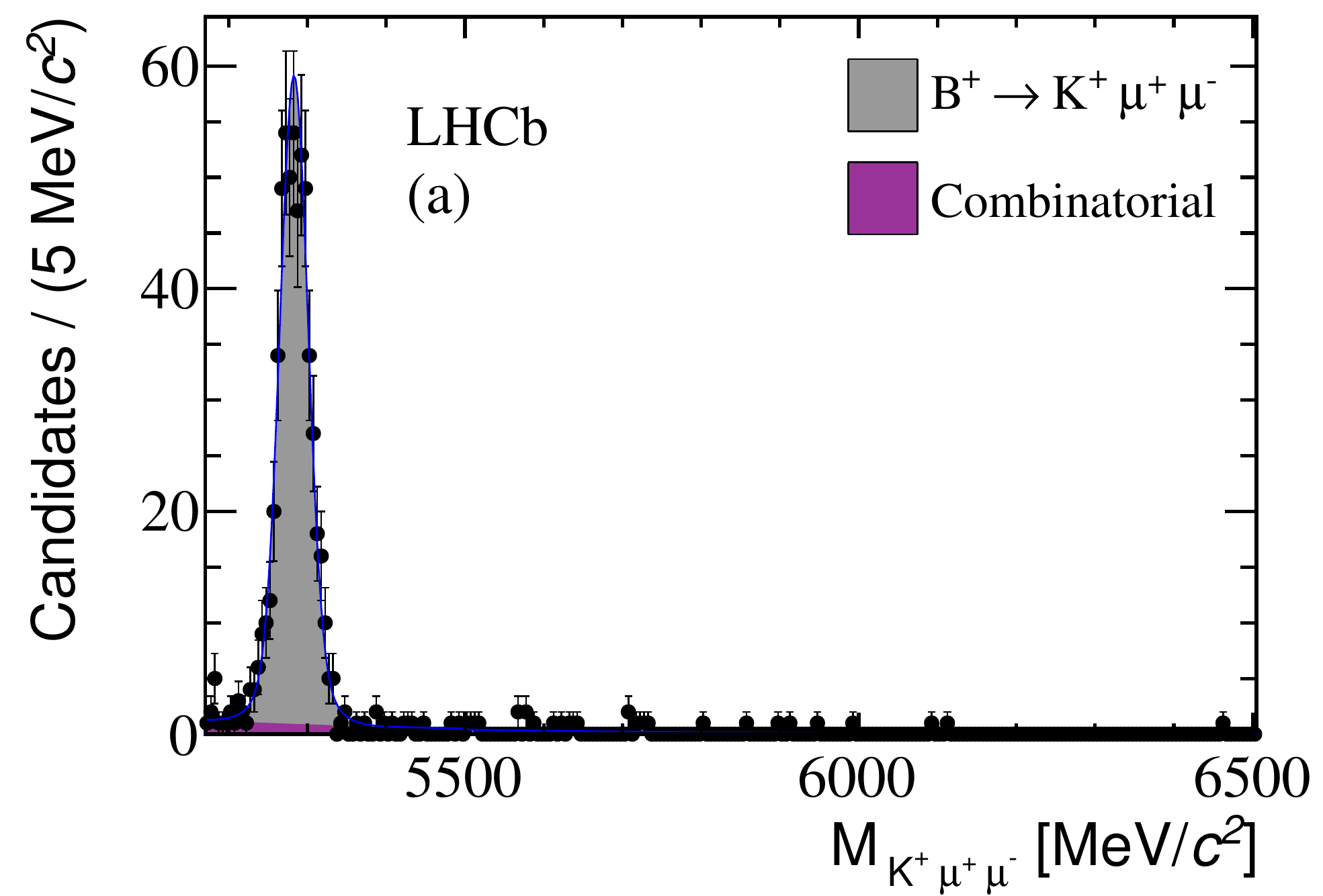}
   \includegraphics[width=0.49\textwidth] {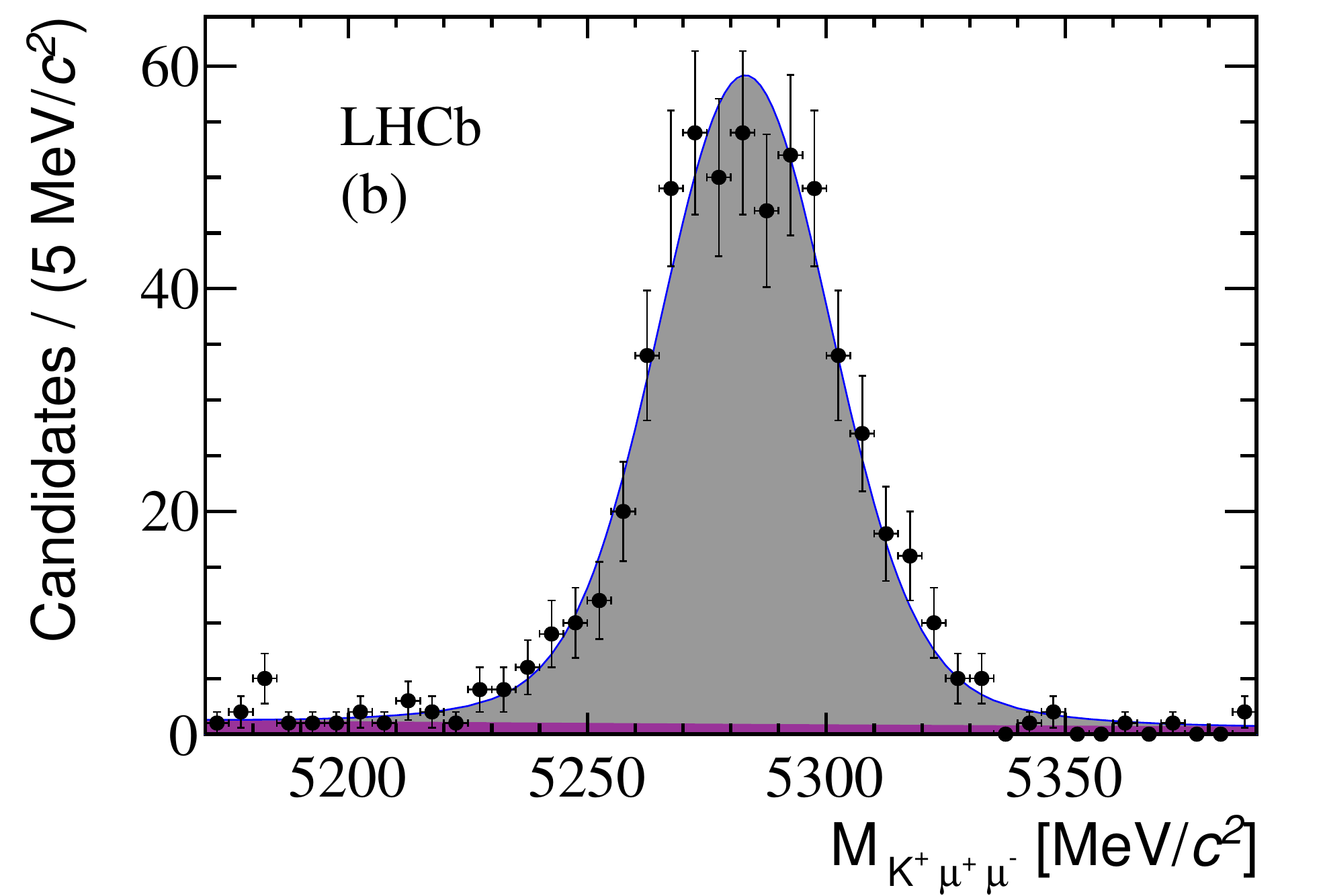}

  \caption{Invariant mass distribution of \kmumu candidates with the fit projection overlaid (a) in the full mass range and (b) in the region around the \B mass. In the legend, ``combinatorial'' refers to the combinatorial background. 
  }
   \label{fig:kmumu}
\end{figure}

The $M_{\pip \mup \mun}$ and $M_{\Kp \mup \mun}$ distributions are shown in Figs~\ref{fig:pimumu} and~\ref{fig:kmumu}, respectively. The fit gives a \pimumu signal yield of 25.3~$^{+6.7}_{-6.4}$, and a \kmumu signal yield of 553~$^{+24}_{-25}$.

\subsection{Cross check of the fit procedure}
\label{sec:fit:jpsipi}
\begin{figure}
  \centering
   \includegraphics[width=0.6\textwidth] {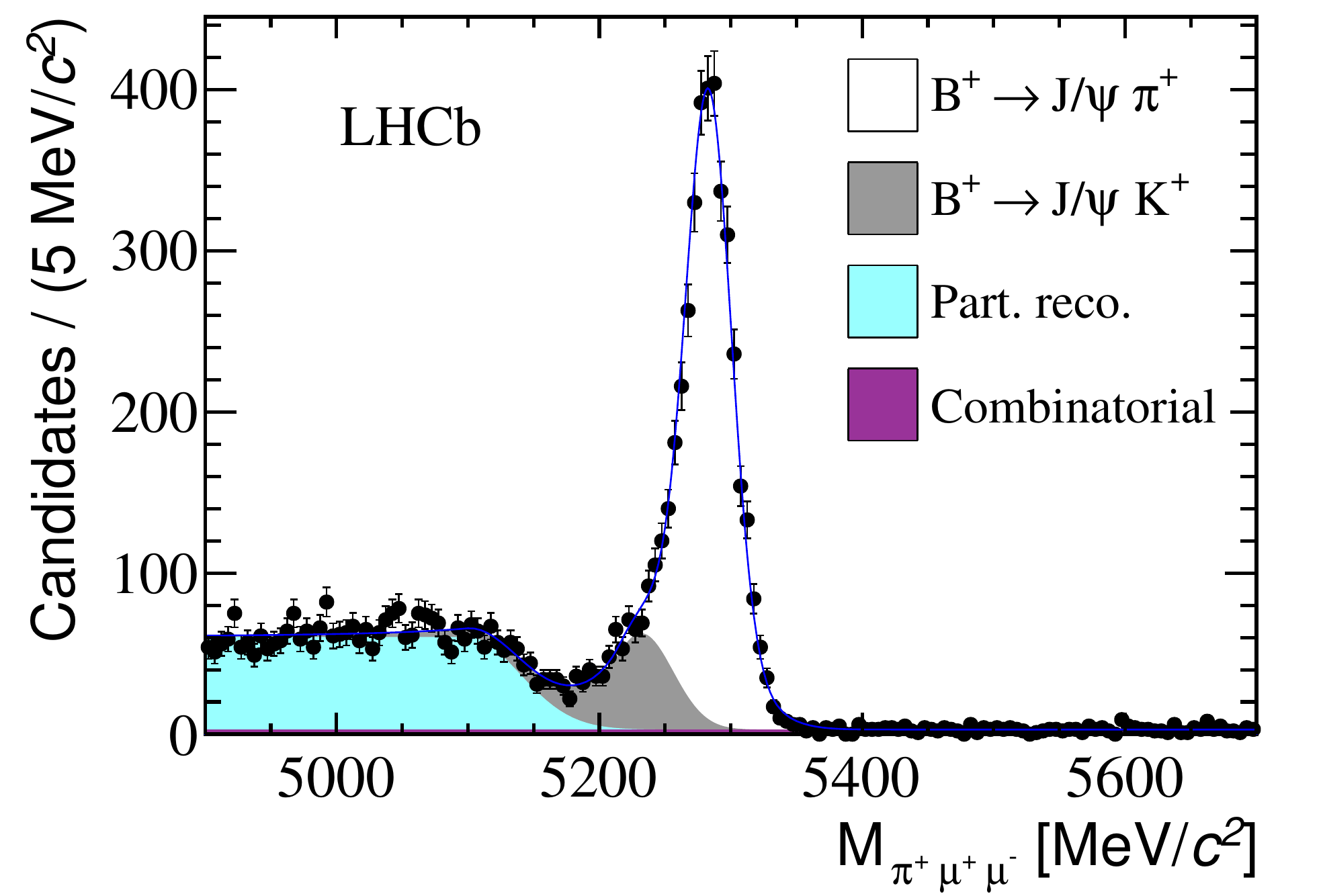}
   \label{fig:jpsipifit}
  \caption{Invariant mass distribution of ${B^+\to J\!/\!\psi \pip}$ candidates with the fit projection overlaid. In the legend, ``part. reco.'' and ``combinatorial'' refer to partially reconstructed and combinatorial backgrounds respectively. 
The fit model is described in the text.}
\end{figure}

The fit procedure was cross-checked on \jpsipi decays, accounting for the background from \jpsik decays.
The resulting fit is shown in Fig.~\ref{fig:jpsipifit}. The shape of the combined ${B^+\to J\!/\!\psi \pip}$ and ${B^+\to J\!/\!\psi K^+}$ mass distribution is well reproduced. 
The ${B^+\to J\!/\!\psi K^+}$ yield is not constrained in this fit. The fitted yield of 1024 $\pm$ 61 candidates is consistent with the expectation of 958 $\pm$ 31~\stat candidates. This expectation is again computed by weighting the \jpsik candidates, which are isolated using a kaon PID requirement, according to the PID efficiency derived from $D^{*+} \to (D^{0} \to \Km \pip) \pip$ events.

\section{Determination of branching fractions}
\label{sec:norm}

The \pimumu branching fraction is given by 
\begin{eqnarray}
\BR(\pimumu) &=& \frac{\BR({\jpsik})}{N_{\jpsik}} {\frac{\epsilon_{{B^+\to J\!/\!\psi K^+}}}{{\epsilon_{\decay{\Bp}{\pip \mup\mun}}}}} N_{\pimumu}\\
&=& \alpha \cdot N_{\pimumu}\,,
\end{eqnarray}
where \BR$(X)$, $N_{X}$ and $\epsilon_{X}$ are the branching fraction, the number of events and the total efficiency, respectively, for decay mode $X$, and $\alpha$ is the single event sensitivity. The total efficiency includes reconstruction, trigger and selection efficiencies.
The ratio $\epsilon_{\jpsik}/\epsilon_{\pimumu}$  is determined to be 1.60 $\pm$ 0.01 using simulated events, where the uncertainty is due to the limited sizes of the simulated samples only. Other sources of systematic uncertainty are discussed in Sect.~\ref{sec:Systematic}.
The difference in efficiencies between \mbox{{\decay{\Bp}{\pip \mup\mun}}} and \mbox{${B^+\to J\!/\!\psi K^+}$} events is largely due to the mass vetoes used to remove the charmonium resonances, and the different PID requirements. 
The \jpsikmm branching fraction is \mbox{(6.02 $\pm$ 0.20)$\times 10^{-5}$~\cite{PDG2012}}. 
Together with the other quantities in Eq.~2, this gives a single event sensitivity of \mbox{$\alpha$ = (9.1~$\pm $ 0.1)$\times 10^{-10}$}, where the uncertainty is due to the limited sizes of the simulated samples only.

The ratio of \pimumu and \kmumu branching fractions is given by
 \begin{equation}
R =  \frac{N_{\pimumu}}{N_{\kmumu}} \frac{\epsilon_{\kmumu}}{\epsilon_{\pimumu}} \,,
\end{equation}
where simulated events give ${\epsilon_{{\kmumu}}}/{\epsilon_{\pimumu}}$ = 1.15~$\pm$~0.01.

\section{Systematic uncertainties}
\label{sec:Systematic}

Two sources of systematic uncertainties are considered: those affecting the determination of the \pimumu and \kmumu signal yields, and those affecting only the normalisation.

Uncertainties in the shape parameters for the misidentified \kmumu PDF in the fit are taken into account by including Gaussian constraints on their values. The most significant sources of uncertainty  in the determination of these shape parameters arise from the procedure for correcting the \jpsik mass shape to match that of the \kmumu decay, and the correction for the hadron PID requirements. 
The uncertainty on the \pimumu yield determined with the fit takes these shape parameter uncertainties into account, and they are therefore included in the statistical rather than the systematic uncertainty. These uncertainties affect the \pimumu yield at below the one percent level.
None of these effects give rise to any significant uncertainty for the \kmumu decay.

Uncertainties on the two efficiency ratios $\epsilon_{\jpsik}/\epsilon_{\pimumu}$ and ${\epsilon_{{\kmumu}}}/{\epsilon_{\pimumu}}$ affect the conversion of the \pimumu yield into a branching fraction, and the measurement of the ratio of branching fractions \kpifrac.
The largest systematic uncertainty on these efficiency ratios is the choice of form factors used to generate the simulated events. Using an alternative set of form factors changes the \pimumu efficiency by 3\%, and this difference is taken as a systematic uncertainty. For the ratio of \pimumu and \kmumu, the alternative form factors are used for both \pimumu and \kmumu, giving a systematic uncertainty of 1.7\%.
To estimate the uncertainty arising from the PID efficiency,  the ratio of corrected yields between the \jpsik and \jpsipi decay modes is measured, varying the PID requirements. The largest resulting difference 
with respect to the nominal value is 1.1\%, which is taken as the systematic uncertainty. 

The systematic uncertainty arising from the knowledge of the trigger efficiency is determined using \jpsik candidates in the data. Taking the events which pass the trigger independently of the \jpsik candidate, the fraction of these events which also pass the trigger based on the \jpsik candidate provides a determination of the trigger efficiency. 
The efficiency determined in this way is compared to that calculated in simulated events using the same method, and the difference is taken as the systematic uncertainty. This gives a 1.4\% uncertainty on $\epsilon_{\jpsik}/\epsilon_{\pimumu}$ and ${\epsilon_{{\kmumu}}}/{\epsilon_{\pimumu}}$.

For all decays under consideration, there are small differences between the distributions of some reconstructed quantities in the data and in the simulated events. These differences are assessed by comparing the distributions of data and simulated events for \jpsik candidates. The 
simulation is corrected to match the data where it disagrees, and the resulting 0.4\% difference between the raw and corrected ratio of \jpsik and \pimumu efficiencies is taken as a systematic uncertainty. 
The statistical uncertainty from the limited simulation sample size is 0.7\%. When normalising to \jpsik, the measured \jpsik and \Jpsimumu branching fractions contribute an uncertainty of 3.5\% to the \pimumu branching fraction. 
The systematic uncertainties are summarised in Table~\ref{tab:syst}. 

\begin{table}
\caption{Summary of systematic uncertainties.}
  \centering
  \begin{tabular}{ l | c | c}
	Source & \BR(\pimumu) (\%) & $\frac{\BR(\pimumu)}{\BR(\kmumu)}$ (\%)\\
      \hline
        Form factors   & 3.0 & 1.7 \\
	Trigger efficiency & 1.4 & 1.4 \\
      	PID performance & 1.1 & 1.1 \\
	Data simulation differences & 0.4 & 0.4\\
	Simulation sample size & 0.7 & 0.7 \\
	\BR(\jpsikmm) & 3.5 & -- \\
	\hline 
	Total & 5.0 & 2.6 \\
    \end{tabular}
     \label{tab:syst}
\end{table}

\section{Results and conclusion}
\label{sec:results}

The statistical significance of the \pimumu signal observed in Fig.~\ref{fig:pimumu} is computed from the difference in the minimum log-likelihood between the signal-plus-background and background-only hypotheses. Both the statistical and systematic uncertainties on the shape parameters (which affect the significance) are taken into account. 
The fitted yield corresponds to an observation of the \pimumu decay with 5.2~$\sigma$ significance.
This is the first observation of 
a \dellell transition. 
Normalising the observed signal to the \mbox{\jpsik} decay, using the single event sensitivity given in Sect.~\ref{sec:norm},
the branching fraction of the \mbox{\pimumu} decay is measured to be
\begin{equation*}
 \BR(\pimumu) = (2.3~\pm~0.6~\stat~\pm~0.1~\syst)\times 10^{-8}\,.
\end{equation*}
This is compatible with the SM expectation of \mbox{(2.0 $\pm$ 0.2)$\times 10^{-8}$~\cite{Wang:2007sp}}. Given the agreement between the present measurement and the SM
prediction, contributions from physics beyond the SM can only modify the \pimumu branching fraction by a small amount. A significant
improvement in the precision of both the experimental measurements and
the theoretical prediction will therefore be required to resolve any new
physics contributions.

Taking the measured \kmumu yield and ${\epsilon_{{\kmumu}}}/{\epsilon_{\pimumu}}$, the ratio of \mbox{\pimumu} and \mbox{\kmumu} branching fractions is measured to be
\begin{equation*}
 \frac{\BR(\pimumu)}{\BR(\kmumu)} = 0.053~\pm~0.014~\stat~\pm~0.001~\syst\,.
\end{equation*}
In order to extract \vtdts~from this ratio of branching fractions, the SM expectation for the ratio of \pimumu and \kmumu branching fractions is calculated using the \evtgen package~\cite{Lange:2001uf}, which implements the calculation in Ref.~\cite{Ali:1999mm}. This calculation has been updated with the expressions for Wilson coefficients and power corrections from Ref.~\cite{Ali:2002jg}, and formulae for the $q^{2}$ dependence of these coefficients from Refs.~\cite{Asatrian:2001de,Bobeth:1999mk}. 
Using this calculation, and form factors taken from Ref.~\cite{Ball:2004ye} (``set II''), the integrated ratio of form factors and Wilson coefficients is determined to be $f = 0.87$.
Neglecting theoretical uncertainties, the measured ratio of \pimumu and \kmumu branching fractions then gives
\begin{equation*}
|V_{\text{td}}| / |V_{\text{ts}}| = \frac{1}{f} \sqrt{\frac{\BR(\pimumu)}{\BR(\kmumu)}} =0.266~\pm~0.035~\stat~\pm~0.003~\syst,
\end{equation*}
which is compatible with previous determinations~\cite{Amhis:2012bh,delAmoSanchez:2010ae,Abe:2005rj,Aubert:2006pu}. An additional uncertainty will arise from the knowledge of the form factors. As an estimate of the scale of this uncertainty, the ``set~IV'' parameters available in Ref.~\cite{Ball:2004ye} change the value of $|V_{\text{td}}| / |V_{\text{ts}}|$ by 5.1\%. This estimate is unlikely to cover a one sigma range on the form factor uncertainty, and does not take into account additional sources of uncertainty beyond the form factors. A full theoretical calculation taking into account such additional uncertainties, which also accurately determines the uncertainty on the ratio of form factors, would allow a determination of \vtdts~with comparable precision to that from radiative penguin decays.

\section*{Acknowledgements}

\noindent We express our gratitude to our colleagues in the CERN accelerator
departments for the excellent performance of the LHC. We thank the
technical and administrative staff at CERN and at the LHCb institutes,
and acknowledge support from the National Agencies: CAPES, CNPq,
FAPERJ and FINEP (Brazil); CERN; NSFC (China); CNRS/IN2P3 (France);
BMBF, DFG, HGF and MPG (Germany); SFI (Ireland); INFN (Italy); FOM and
NWO (The Netherlands); SCSR (Poland); ANCS (Romania); MinES of Russia and
Rosatom (Russia); MICINN, XuntaGal and GENCAT (Spain); SNSF and SER
(Switzerland); NAS Ukraine (Ukraine); STFC (United Kingdom); NSF
(USA). We also acknowledge the support received from the ERC under FP7
and the Region Auvergne.

\bibliographystyle{LHCb}
\bibliography{main}

\end{document}